# Four-dimensional Cone Beam CT Reconstruction and Enhancement using a Temporal Non-Local Means Method

**Xun Jia[1], Zhen Tian[1], Yifei Lou[2], Jan-Jakob Sonke[3], and Steve B. Jiang[1]**

[1]Center for Advanced Radiotherapy Technologies and Department of Radiation Medicine and Applied Sciences, University of California San Diego, La Jolla, CA 92037, USA
[2]School of Electrical and Computer Engineering, Georgia Institute of Technology, Atlanta, GA 30318, USA
[3]Department of Radiation Oncology, The Netherlands Cancer Institute-Antoni van Leeuwenhoek Hospital, Plesmanlaan 121, 1066 CX Amsterdam, The Netherlands

E-mail: sbjiang@ucsd.edu, xujia@ucsd.edu

**Purpose**: Four-dimensional Cone Beam Computed Tomography (4D-CBCT) has been developed to provide respiratory phase resolved volumetric imaging in image guided radiation therapy (IGRT). Conventionally, it is reconstructed by first sorting the x-ray projections into multiple respiratory phase bins according to a breathing signal extracted either from the projection images or some external surrogates, and then reconstructing a 3D CBCT image in each phase bin independently using FDK algorithm. This method requires adequate number of projections for each phase, which can be achieved using a low gantry rotation or multiple gantry rotations. Inadequate number of projections in each phase bin results in low quality 4D-CBCT images with obvious streaking artifacts. 4D-CBCT images at different breathing phases share a lot of redundant information, because they represent the same anatomy captured at slightly different temporal points. Taking this redundancy along the temporal dimension into account can in principle facilitate the reconstruction in the situation of inadequate number of projection images. In this work, we propose two novel 4D-CBCT algorithms: an iterative reconstruction algorithm and an enhancement algorithm, utilizing a temporal nonlocal means (TNLM) method. **Methods:** We define a TNLM energy term for a given set of 4D-CBCT images. Minimization of this term favors those 4D-CBCT images such that any anatomical features at one spatial point at one phase can be found in a nearby spatial point at neighboring phases. 4D-CBCT reconstruction is achieved by minimizing a total energy containing a data fidelity term and the TNLM energy term. As for the image enhancement, 4D-CBCT images generated by the FDK algorithm are enhanced by minimizing the TNLM function while keeping the enhanced images close to the FDK results. A forward-backward splitting algorithm and a Gauss-Jacobi iteration method are employed to solve the problems. The algorithms are implemented on GPU to achieve a high computational efficiency. Our algorithms have been tested on a digital NCAT phantom and a clinical patient case. **Results:** The reconstruction algorithm and the enhancement algorithm generate visually similar 4D-CBCT images, both better than the FDK results. Quantitative evaluations indicate that, compared with the FDK results, our reconstruction method improves contrast-to-noise-ratio (CNR) by a factor of 2.56~3.13 and our enhancement method increases the CNR by 2.75~3.33 times. The enhancement method also removes over 80% of the streak artifacts from the FDK results. The total computation time is ~460 sec for the reconstruction algorithm and ~610 sec for the enhancement algorithm on an NVIDIA Tesla C1060 GPU card. **Conclusions:** By innovatively taking the temporal redundancy among 4D-CBCT images into consideration, the proposed algorithms can produce high quality 4D-CBCT images with much less streak artifacts than the FDK results, in the situation of inadequate number of projections.

## 1. Introduction

Cone Beam Computed Tomography (CBCT) has been widely used in image guided radiation therapy (IGRT) to provide fast and accurate 3D image guidance prior to a radiation treatment[1, 2]. However, when CBCT is applied to thorax or upper abdomen regions, the image quality can be heavily degraded due to patient respiratory motion. In particular, serious motion-induced artifacts, such as blurring or distortion, can be observed, which potentially compromises the efficacy of CBCT in IGRT. To overcome this problem, four-dimensional CBCT (4D-CBCT), or respiratory correlated CBCT[3-7], has been developed to provide respiratory phase-resolved volumetric images for IGRT. In such an imaging modality, all the x-ray projections are first retrospectively grouped into different respiratory phase bins according to a breathing signal tagged on every projection image. A CBCT image for each breathing phase is then reconstructed independently. This procedure of binning x-ray projections considerably relieves the data inconsistency problem among projections, yielding CBCT images with much less motion-induced artifacts. More importantly, this approach provides us a set of phase-resolved volumetric images, which are of particular use when treating tumors inside organs with appreciable motion, such as lung.

Though 4D-CBCT is capable of reducing the motion artifacts, it poses another challenge. In fact, the phase binning approach leads to insufficient number of x-ray projections in each respiratory phase bin and thus causes severe streaking artifacts, when a standard 3D-CBCT scanning protocol is applied. In the past, many attempts have been made toward removing or relieving this problem. For example, scanning protocols of multiple gantry rotations and slow gantry rotations[5, 7, 8] have been proposed to considerably increase the projection number per phase. Nonetheless, the long data acquisition process makes them not practically preferable. The reduced mAs levels to avoid amplified imaging dose to a patient also decrease signal-to-noise ratio and hence degrade image quality. A number of research efforts have also been made to achieve satisfactory image quality by post-processing the 4D-CBCT images via advanced image enhancement algorithms. For instance, a prior image based approach[9] has been developed by first reconstructing a blurred CBCT images with all projections and then using it to estimate and remove the streaking artifacts from the 4D-CBCT images at various phases. It has also been proposed to enhance the CBCT image by first deforming images at all phases into a single one and superimposing them together[10]. The efficacy of these approaches, however, largely depends on the accuracy of the algorithms involved, such as the deformable image registration algorithm.

Recently, nonlocal operators have become an effective tool for solving image restoration problems[11-13], where similar structures in a single image to be restored are identified and grouped to constructively enhance each other. The underlying assumption is that, the image to be recovered contains repetitive features. This concept can be extended to the temporal domain. In the context of 4D-CBCT, repetitive features can be found in the images of neighboring phases, with slightly varied locations and appearances. In other words, 4D-CBCT images are highly redundant along the temporal dimension. In principle, taking this temporal redundancy into account can greatly facilitate the 4D-CBCT reconstruction in the situation of insufficient number of projections. Inspired by this idea, we have previously studied the 4D-CT reconstruction problem by proposing a generalized non-local means method, temporal non-local means (TNLM)[14, 15]. In this paper, we will present our recent work on solving a 4D-CBCT reconstruction problem using the TNLM method. In particular, we will reconstruct the 4D-CBCT images using an iterative algorithm derived from the minimization of an energy function containing a TNLM form and a data fidelity term. At each iteration step, similar anatomical features in successive phases are identified and averaged. This algorithm is found to be effective to remove artifacts while preserving true patient anatomy.

One disadvantage of the TNLM method is its inherent high computational cost. It will take a long time to perform the reconstruction of a set of 4D-CBCT images with, for example, 10 phases, since the searching for similar features has to be invoked in each iteration and for each voxel. To partially relieve this problem, we will also propose a TNLM-based 4D-CBCT enhancement algorithm in this paper. As such, 4D-CBCT images are first reconstructed by the conventional FDK algorithm[16]. Post-processing on the 4D-CBCT image is performed by utilizing an operator derived from the TNLM method. This method is in particular effective to remove the streaking artifacts caused by the FDK algorithm when reconstructing a 4D-CBCT image with insufficient projections at each phase.

Another approach to shorten the computation time is to use a graphics processing unit (GPU) platform. Recently, the rapid development of GPU technology for scientific computing has offered us a promising prospect to speed up computationally heavy tasks in radiotherapy, such as CBCT reconstruction[17-22], deformable image registration[18, 23, 24], dose calculation[25-30], treatment plan optimization[31-33], and many other tasks[34]. In this work, we have also implemented both the reconstruction and the enhancement algorithms on a GPU platform and a high computational efficiency has been observed.

## 2. Methods

### *2.1 4D-CBCT Reconstruction Model*

Let us divide a respiratory cycle into $N_p$ phases labeled by $i = 1,2,\ldots,N_p$. The 4D-CBCT image of phase $i$ is denoted by a vector $f_i$. $P_i$ is the projection matrix of phase $i$ that maps the image $f_i$ into a set of projections corresponding to various projection angles. The measured projections for this phase are denoted by a vector $y_i$. We attempt to reconstruct the 4D-CBCT images by solving the following optimization problem:

$$\{f_i(x)\} = \mathrm{argmin}_{\{f_i\}} \sum_{i=1}^{N_p} \left\{\frac{\mu}{2}\|P_i f_i - y_i\|_2^2 + \frac{1}{2}J[f_i, f_{i+1}]\right\}, \quad (1)$$

where the first term in the summation is a data fidelity term, ensuring that the projections of the reconstructed 4D-CBCT images at each phase match the corresponding observations. $\|.\|_2$ stands for the standard $l_2$ norm of a vector. The second term, $J[\cdot,\cdot]$ is the regularization term imposed on neighboring phases as a temporal regularization to explore the inter-phase redundancy. A constant $\mu > 0$ in Eq. (1) adjusts the relative weight between the data fidelity and regularization terms. A periodic boundary condition along the temporal direction is assumed, *i.e.*, $f_{N_p+1} = f_1$. Note that, in this approach, we are reconstructing the images at all phases $\{f_i\}$ altogether instead of reconstructing each of them independently.

As for the regularization term, we use a recently proposed TNLM function[14, 15] to impose regularizations along the temporal direction between 4D-CBCT images at successive respiratory phases. As such, for two volumetric images $f_i$ and $f_j$, $J[f_i, f_j]$ is defined as:

$$J[f_i, f_j] = \iint \mathrm{d}\boldsymbol{x}\mathrm{d}\boldsymbol{y}\, \left[f_i(\boldsymbol{x}) - f_j(\boldsymbol{y})\right]^2 w_{f_i^*, f_j^*}(\boldsymbol{x}, \boldsymbol{y}), \tag{2}$$

where $\boldsymbol{x}$ and $\boldsymbol{y}$ are coordinates on the image $i$ and $j$, respectively. Suppose we know the ground truth images $f_i^*(\boldsymbol{x})$ and $f_j^*(\boldsymbol{x})$, the weighting factors $w_{f_i^*, f_j^*}(\boldsymbol{x}, \boldsymbol{y})$ are defined based on them and are independent of $f_i(\boldsymbol{x})$ and $f_j(\boldsymbol{x})$. Specifically,

$$w_{f_i^*, f_j^*}(\boldsymbol{x}, \boldsymbol{y}) = \frac{1}{Z} \exp\left[-\frac{1}{2h^2} \left\| p_{f_i^*}(\boldsymbol{x}) - p_{f_j^*}(\boldsymbol{y}) \right\|_2^2\right], \tag{3}$$

where $p_{f_i^*}(\boldsymbol{x})$ denotes a small cubic volume in the image $f_i^*$ centering at the coordinate $\boldsymbol{x}$. $Z$ is a normalization factor such that $\int \mathrm{d}\boldsymbol{x}\, w_{f_i^*, f_j^*}(\boldsymbol{x}, \boldsymbol{y}) = 1$. Yet, since we would never know the solution before performing the reconstruction, we will estimate the weighting factors during the reconstruction process using the latest available solutions, as will be described in subsequent Sections. This TNLM regularization term compares every pair of voxels, namely $\boldsymbol{x}$ in image $f_i$ and $\boldsymbol{y}$ in image $f_j$. If they are considered similar, a relatively high weighting factor will be assigned to this pair. The similarity is quantified by computing the $l_2$ distance between the two cubic volumes centered at those two voxels. The scalar $h$ controls to what extent the similarity between volumes is enforced. The underlying reason why such a TNLM term will impose inter-phase similarity will be discussed in Section 2.3.

### *2.2 4D-CBCT Enhancement Model*

In this paper, we also propose an image enhancement model to directly improve the image quality of those 4D-CBCT images reconstructed from available algorithms, such as the conventional FDK-type algorithms[16], using the phase binned x-ray projections. Let us consider a set of 4D-CBCT images $\{g_i(\boldsymbol{x})\}_{i=1}^{N_p}$. Due to the insufficient number of projections in each phase bin, serious streaking artifacts are expected in $\{g_i(\boldsymbol{x})\}$. It is the objective of this enhancement model to directly remove these streaks and produce a new set of 4D-CBCT images $\{f_i(\boldsymbol{x})\}_{i=1}^{N_p}$ by exploiting the temporal redundancy between images at successive phases. This goal can be achieved by solving the following optimization problem:

$$\{f_i(\boldsymbol{x})\} = \mathrm{argmin}_{\{f_i\}} \left\{\frac{\mu}{2} \sum_i \|f_i - g_i\|_2^2 + \frac{1}{2} J[f_i, f_{i+1}]\right\}. \tag{4}$$

The first term ensures that the enhanced images do not largely deviate from the input low quality images, while the second term imposes the temporal regularization conditions on the solution.

It is worth mentioning that in both models, we exclude the nonlocal regularization terms $[f_i, f_i]$ from the energy function, namely those terms comparing cubic volumes within a single phase image, for the following two considerations. First, the efficacy of TNLM approaches relies on the fact that similar feathers at different spatial-temporal locations can be utilized to constructively enhance each other. It is expected that similar features can exist at spatially different locations in two neighboring phases due to the smooth respiratory motion. Yet, similar structures are hardly found in a CBCT image at a given phase. Second, it is our main goal to remove streaking artifacts caused by the insufficient number of projections in each breathing phase. If the regularization term $J[f_i, f_i]$ were used, the streaking artifacts would be in fact strengthened rather than suppressed, since this term tends to locate those straight lines in a single image and consider them to be similar to each other. On the other hand, since the x-ray projections are usually along different directions at two different breathing phases, the streaking artifacts do not repeat themselves at different phases and thus will be suppressed the TNLM term proposed in this paper.

*2.3 Algorithms*

Let us first present the algorithm for solving the reconstruction problem posed by the Eq. (1). We utilize a forward-backward splitting algorithm[35, 36], which allows us to obtain the solution by iteratively solving the following two sub-problems:

$$\begin{aligned} &\text{(P1): } \left\{g_i^{(k)}\right\} = \text{argmin}_{\{g_i\}} \textstyle\sum_{i=1}^{N_p} \|P_i g_i - y_i\|_2^2, \\ &\text{(P2): } \left\{f_i^{(k)}\right\} = \text{argmin}_{\{f_i\}} \textstyle\sum_{i=1}^{N_p} \left\{\frac{\mu}{2}\left\|f_i - g_i^{(k)}\right\|_2^2 + \frac{1}{2} J[f_i, f_{i+1}]\right\}, \end{aligned} \tag{5}$$

where $k$ is the index for iteration steps. In particular, (P1) itself is an optimization problem and is usually solved by an iterative algorithm. Due to the underdetermined nature of this subproblem, its solution depends on the initial value. In practice, this initial value at iteration $k$ is taken to be $\left\{f_i^{(k-1)}\right\}$.

The energy function in the sub-problem (P1) in Eq. (5) is of a simple quadratic form. This problem can therefore be easily solved using a gradient-based minimization approach. For the consideration of efficiency, conjugate gradient least square (CGLS) method [37] is used. Since the images at different phases are uncoupled in this sub-problem, this minimization problem can actually be solved in a phase-by-phase manner.

Note that the intermediate variables $g_i^{(k)}$ are obtained purely based on the data fidelity condition, it is expected that it contains anatomical structures but contaminated by serious artifacts like streaks. The purpose of the subsequent sub-problem (P2) is to remove those artifacts while preserving the true structures using the inter-phase similarity. To solve the sub-problem (P2) of Eq. (5), let us first take functional variation of $E[f_i; g_i] = \frac{\mu}{2}\sum_i \|f_i -$ $g_i\|_2^2 + \frac{1}{2}J[f_i, f_{i+1}]$ with respect to $f_i(\boldsymbol{x})$. The iteration index $k$ on $g_i$ has been omitted here to simplify notation. Note that the weighting factors $w_{f_i^*,f_j^*}(\boldsymbol{x},\boldsymbol{y})$ are constants defined according to the ground truth images $f_i^*(\boldsymbol{x})$ and $f_j^*(\boldsymbol{x})$. We arrive at

$$\frac{\delta E}{\delta f_i(\boldsymbol{x})} = \mu(f_i - g_i) + 2f_i - \int \mathrm{d}\boldsymbol{y}\, f_{i+1}(\boldsymbol{y}) w_{f_i^*,f_{i+1}^*}(\boldsymbol{x},\boldsymbol{y}) - \int \mathrm{d}\boldsymbol{y}\, f_{i-1}(\boldsymbol{y}) w_{f_i^*,f_{i-1}^*}(\boldsymbol{x},\boldsymbol{y})\ . \tag{6}$$

By setting this functional variation zero, we obtain the optimality condition for the minimization of $E[f_i; g_i]$. If we rewrite this optimality condition as

$$f_i(\boldsymbol{x}) = \frac{\mu}{2+\mu} g_i + \frac{1}{2+\mu}\left[\int \mathrm{d}\boldsymbol{y}\, f_{i+1}(\boldsymbol{y}) w_{f_i^*,f_{i+1}^*}(\boldsymbol{x},\boldsymbol{y}) + \int \mathrm{d}\boldsymbol{y}\, f_{i-1}(\boldsymbol{y}) w_{f_i^*,f_{i-1}^*}(\boldsymbol{x},\boldsymbol{y})\right], \tag{7}$$

we can construct a Gauss-Jacobi type iterative scheme[38] for solving the equation posed by the optimality condition as

$$f_i^{(l+1)}(\boldsymbol{x}) = \frac{\mu}{2+\mu} g_i(\boldsymbol{x}) + \frac{1}{2+\mu}\left[\int \mathrm{d}\boldsymbol{y}\, f_{i+1}^{(l)}(\boldsymbol{y}) w_{f_i^*,f_{i+1}^*}(\boldsymbol{x},\boldsymbol{y}) + \int \mathrm{d}\boldsymbol{y}\, f_{i-1}^{(l)}(\boldsymbol{y}) w_{f_i^*,f_{i-1}^*}(\boldsymbol{x},\boldsymbol{y})\right], \tag{8}$$

where $l$ is the iteration index for this subproblem. It follows from Theorem 10.1.1 in Golub (1996)[38] that such an iteration scheme converges for any $\mu > 0$. Note that our reconstruction algorithm iteratively solves the two sub-problems (P1) and (P2) and this iterative scheme in Eq. (8) is used for solving (P2), which is to be performed a number of times during the entire iteration, it is not necessary to solve (P2) accurately. In practice, we only perform (8) once each time (P2) is solved.

The meaning of Eq. (8) is straightforward. At each iteration step, the algorithm updates the solution to $f_i^{(l+1)}$ via a weighted average over the image $g_i$ and the images at neighboring phases $f_{i+1}^{(l)}$ and $f_{i-1}^{(l)}$. In particular, as in the square bracket in Eq. (8), this update incorporates information from images at neighboring phases in a nonlocal fashion. As such, any features that repetitively appear in successive phases, such as true anatomical structures, are preserved during the iteration. In contrast, those features do not repeat, such as streaking artifacts, are suppressed.

Moreover, since the weighting factors $w_{f_i^*,f_j^*}(\boldsymbol{x},\boldsymbol{y})$ are defined according to the ground truth images $f_i^*(\boldsymbol{x})$ and $f_j^*(\boldsymbol{y})$ that are not know beforehand, we estimate these weights during the iteration according to the latest available images $g_i^{(l)}(\boldsymbol{x})$ and $g_j^{(l)}(\boldsymbol{y})$ as

$$w_{f_i^*,f_j^*}(\boldsymbol{x},\boldsymbol{y}) \approx \frac{1}{Z}\exp\left[-\frac{1}{2h^2}\left\|p_{g_i^{(k)}}(\boldsymbol{x}) - p_{g_j^{(k)}}(\boldsymbol{y})\right\|_2^2\right]. \tag{9}$$

Since a reconstructed 4D-CBCT image physically represents the x-ray attenuation coefficient at a spatial point**,** its positiveness has to be ensured during the reconstruction in order to obtain a physically meaningful solution. For this purpose, we perform a correction step on the reconstructed images at each iteration by setting any voxels with negative values to be zero. In practice, we also initialize the reconstruction process by estimating $f_i^{(0)}$ using the FDK algorithm. In summary, the algorithm solving the 4D-CBCT image reconstruction problem is as follows:

**TNLM Reconstruction (TNLM-R) Algorithm:**

Initialize: $f_i^{(0)}$ for $i = 1, \ldots, N_p$.

For $k = 0,1, \ldots$, do the following steps until convergence:

1. Solve (P1) using CGLS with initial value $\{f_i^{(k-1)}\}$ to obtain $\{g_i^{(k)}\}$;
2. Update weights $w_{f_i^*, f_j^*}$ according to Eq. (9) using the image $\{g_i^{(k)}\}$;
3. Compute images $\{f_i^{(k)}\}$ according to Eq. (8);
4. Ensure image positiveness: $f_i^{(k)} = 0$ if $f_i^{(k)} < 0$.

We also note that the enhancement model in Eq. (4) is actually identical to the second sub-problem, (P2) in Eq. (5), in the reconstruction model. Both of them attempt to generate a new image set based on the input $g_i$ by solving the minimization problem. In the enhancement model $g_i$ is the input 4D-CBCT images obtained using another reconstruction algorithm, while in the reconstruction model $g_i$ is an intermediate variable produced by the first sub-problem (P1). As such, the algorithm for the enhancement model is only part of the one for the reconstruction. The only difference is that the weighting factors are estimated using the latest available images $f_i^{(l)}(\boldsymbol{x})$ and $f_j^{(l)}(\boldsymbol{y})$.

**TNLM Enhancement (TNLM-E) Algorithm:**

Initialize: $f_i^{(0)} = g_i$ for $i = 1, \ldots, N_p$.

For $k = 0,1, \ldots$, do the following steps until convergence:

1. Update weight $w_{f_i^*, f_j^*}$ according to Eq. (9) using the image $\{f_i^{(k)}\}$ ;
2. Compute images $\{f_i^{(k+1)}\}$ according to Eq. (8).

*2.4 Implementation*

One drawback of the TNLM-based reconstruction and enhancement algorithms is the high computational burden. During the implementation, a cubic volume $p_{f_i^{(k)}}(\boldsymbol{x})$ centering at the voxel $\boldsymbol{x}$ on the image $f_i$ is compared with cubic volumes centered at all other voxels $\boldsymbol{y}$ on the image $f_j$ to compute the weighting factors $w_{f_i^*, f_j^*}(\boldsymbol{x}, \boldsymbol{y})$. If this cube has $(2d + 1)$ voxels in each dimension, the complexity of such an algorithm is in the order of $O(N^3 N^3 (2d + 1)^3)$, where $N$ is the dimension of the 4D-CBCT images. However, this approach is neither computational efficient nor necessary. In fact, the voxel that is similar to $\boldsymbol{x}$ will locate in its vicinity in the neighboring phases due to the smooth respiratory motion. Therefore, it is adequate to search for the similar voxels only within a search window centering at the voxel $\boldsymbol{x}$ as opposed to searching over the entire image. In practice, we set this search window to be a cubic volume with $(2M+1)$ voxels in each dimension to reduce the algorithmic complexity to $O(N^3(2M+1)^3(2d+1)^3)$. Moreover, since $w_{f_i^*,f_j^*}(\boldsymbol{x},\boldsymbol{y})$ is a mutual weight factor shared by the voxel $\boldsymbol{x}$ and the voxel $\boldsymbol{y}$, apart from the different normalization factors, the nonlocal update step in Eq. (8) can be implemented in such a way that both the image at phase $i$ and the one at phase $j$ are updated simultaneously using the common weighting factor. This strategy can almost save half of the computation time for the nonlocal update step.

To speed up the computation, we also implement our algorithms on NVIDIA CUDA programing environment using an NVIDIA Tesla C1060 card. This GPU card has a total number of 240 processor cores (grouped into 30 multiprocessors with 8 cores each), each with a clock speed of 1.3 GHz. It is also equipped with 4 GB DDR3 memory, shared by all processor cores.

### *2.5 Experiments and metrics*

To evaluate the performance of our algorithms, we have conducted studies on a digital NURBS-based cardiac-torso (NCAT) phantom[39] and one real clinical case. All of the 4D-CBCT images are of a resolution $128^3$ voxels and the voxel size is 0.2 mm along all three spatial dimensions. For the NCAT phantom, it is generated at thorax region with a high level of anatomical realism such as detailed bronchial trees. The virtual patient has a regular respiratory pattern of a period 4 sec and the respiratory cycle is divided into 10 phases. A 2 min 4D-CBCT scan is simulated, within which a total number of 300 x-ray projections equally spaced in a $360^\circ$ gantry rotation are taken. The source-to-isocenter distance and the source-to-detector distance are 100 cm and 153.6 cm, respectively. The x-ray detector size is $40.96\times40.96$ cm$^2$ with a resolution $512\times512$ pixel$^2$. All of these parameters mimic a realistic configuration in Elekta XVI system (Elekta AB, Stockholm, Sweden). The virtual patient is purposely off-center positioned such that the isocenter is on a tumor-like structures inside the lung and the phantom outside the reconstructed region is truncated. For each projection image, we first identify the associated gantry angle and the breathing phase. The x-ray projection is then numerically generated using Siddon's ray-tracing algorithm[40]. These projection images are grouped according to their breathing phases, so that each phase is associated with 30 projections equally spaced in a full $360^\circ$ gantry rotation. The projection angles for the maximum inhale (MI) and the maximum exhale (ME) phases are illustrated in Fig. 1(a).

The real patient is scanned using an Elekta XVI system. A total number of 1169 x-ray projections are acquired in a $200^\circ$ gantry rotation using 4 minutes. The patient is positioned such that the isocenter is inside the tumor in the left lung. The respiratory motion signal is obtained by using Amsterdam Shroud algorithm[41, 42] and the acquired x-ray projections are binned into 10 respiratory phases according to it. Though on average there are 116.9 x-ray projections per phase, this number is only nominal. In fact, the patient breathing cycle is about 4 sec long and only about 60 cycles are covered during the scan. For a given phase at a given breathing cycle, the high x-ray imaging rate leads very similar projection images due to their very close projection angle. Those duplicated projections do not provide substantially different x-ray projection information useful for the 4D-CBCT reconstruction. As a consequence, there are only about 50~60 distinguishable and useful projections in each phase bin. This effect is illustrated in Fig. 1(b).

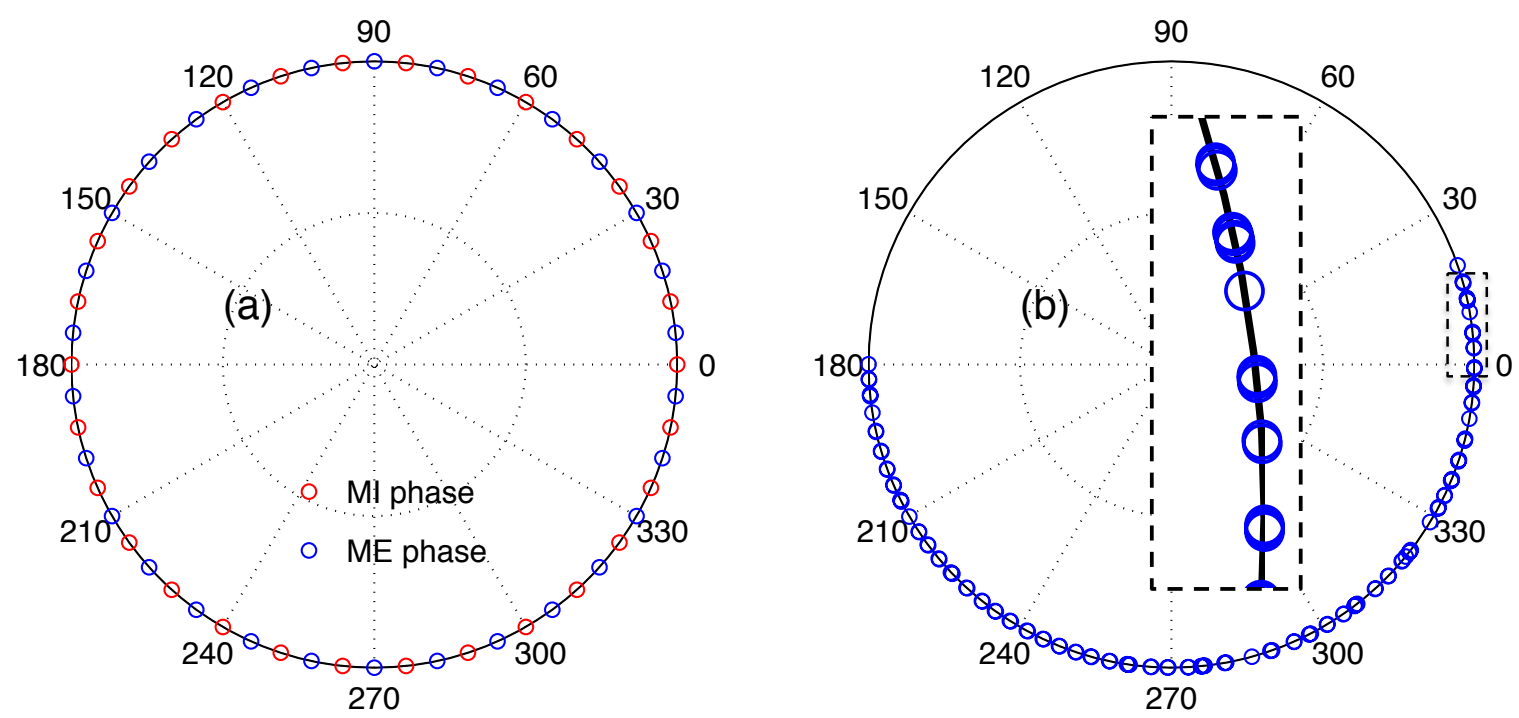

**Figure 1.** Illustration of the x-ray projection angles for (a) the NCAT phantom and (b) the patient case. Each small circle represents one x-ray projection. The insert in (b) shows a zoom in view of the projections around 0 degree to demonstrate the projection clustering issue.

Apart from visual inspections, quantitative metrics are necessary to assess the reconstructed 4D-CBCT image quality. In our studies, the first metrics we utilized is contrast-to-noise ratio ($CNR$). For a given region of interest (ROI), $CNR$ is calculated as $CNR = 2|S - S_b|/(\sigma + \sigma_b)$, where $S$ and $S_b$ are the mean pixel values in the ROI and in a nearby region considered as the background, respectively. $\sigma$ and $\sigma_b$ are the standard deviation of the pixel values inside the ROI and in the background.

The main advantage of our TNLM-based 4D-CBCT enhancement algorithm is its capability of removing streak artifacts from input images. To quantify this effect, we define streak-reduction ratio ($SRR$) to quantitatively measure how much streaks in the input images, namely the FDK results, are removed by the TNLM enhancement algorithm. For a given phase, the $SRR$ is defined as

$$SRR = \frac{TV(f_{FDK}-f^*)-TV(f_{TNLME}-f^*)}{TV(f_{FDK}-f^*)}, \tag{10}$$

where $f^*$ stands for the ground truth images for the corresponding phase. $f_{FDK}$ and $f_{TNLME}$ represent the images reconstructed by the FDK algorithm and our TNLM-E algorithm, respectively. The difference term, such as $(f_{FDK} - f^*)$, is expected to mainly contain the streak artifacts, if there are any. Therefore, by taking a total variation semi-norm defined as $TV(h) = \int \mathrm{d}x|\nabla h(\boldsymbol{x})|$, we are able to use a single number to quantify the amount of streaks in the reconstruction results. Note that this TV term contains spatial image gradient, its value is dominated by the region where the intensity largely fluctuates. As such, $TV(f_{FDK} - f^*) - TV(f_{TNLME} - f^*)$ represents an estimation regarding the absolute amount of streaks that the TNLM-E algorithm removes from the FDK results and hence the $SRR$ defined in Eq. (10) reports this effect in a relative manner. Though the calculation of $SRR$ is straightforward for the NCAT phantom case, one practical difficulty for the patient case is the lack of the ground truth 4D-CBCT images. In practices, we choose $f^*$ to be the CBCT image reconstructed via the FDK algorithm using all the projections at all phases. This is also the image obtained by averaging the ten phases of the FDK results, since FDK reconstruction is a linear operation. The CBCT image chosen as such is free of streaking artifacts due to the large number of projections. However, some anatomical structures in $f^*$ are blurred due to the patient respiratory motion. Though $f^*$ is not the ground truth image any more, it can be considered to be the ground truth image blurred by the motion artifacts. It is expected that using such a $f^*$ can still give us a reasonable estimation regarding the amount of streaks in the reconstructed images. This is because the TV term is only sensitive to the high image gradient parts, and the blurring in $f^*$ will only lead to slowly varying components in the difference images and hence not considerably impacts on the value of the TV term.

## 3. Experimental Results

### *3.1 Visualization of the results*

We first present the reconstructed and the enhanced 4D-CBCT images for the visualization purpose in Fig. 2 and Fig. 3 for the NCAT case and the patient case, respectively. In both figures, the top two rows show the set of 4D-CBCT images reconstructed using the FDK algorithm for a comparison purpose. The middle two rows are the reconstruction results using our TNLM reconstruction algorithm (TNLM-R), while the bottom two rows shows the images obtained from our TNLM enhancement algorithm (TNLM-E) using the FDK results at the top two rows as the input. In each group, transverse, coronal, and sagittal views of the 4D-CBCT images at the MI and ME phases are presented. Due to insufficient number of x-ray projections in each phase, obvious streak artifacts are observed in the 4D-CBCT images reconstructed from the FDK algorithm. On the other hand, the 4D-CBCT images in the other two groups undoubtedly demonstrate the efficacy of the TNLM-R algorithm and the TNLM-E algorithm, as those streaking artifacts are effectively suppressed, while the true anatomical structures, even the tiny ones inside the lung, are well preserved.

### *3.2 Quantitative analysis*

In the two cases we studied, the ROIs for $CNR$ measurements are chosen to be the tumor or a tumor-like structure close to the isocenter in 3 dimensional space, as indicated by the arrows in Fig. 2 and Fig. 3, as these structures are usually of interest in the 4D-CBCT images. We measure the $CNR$s in both cases at each respiratory phase for the results obtained by all the three algorithms, namely FDK, TNLM-R and TNLM-E algorithms. We use the averaged $CNR$s over all 10 phases to represent on average the $CNR$ level for each algorithm for a comparison. The results are shown in Table 1. Due to the insufficient projection numbers in each breathing phase, large streaks lead to high levels of fluctuation of image intensity in the FDK results, causing relatively low $CNR$s for the FDK algorithm. In contrast, both the two TNLM-based algorithms can considerably increase the$CNR$s. These numbers undoubtedly demonstrate that our TNLM-based algorithms outperform the conventional FDK algorithm in terms of contrast-to-noise ratio.

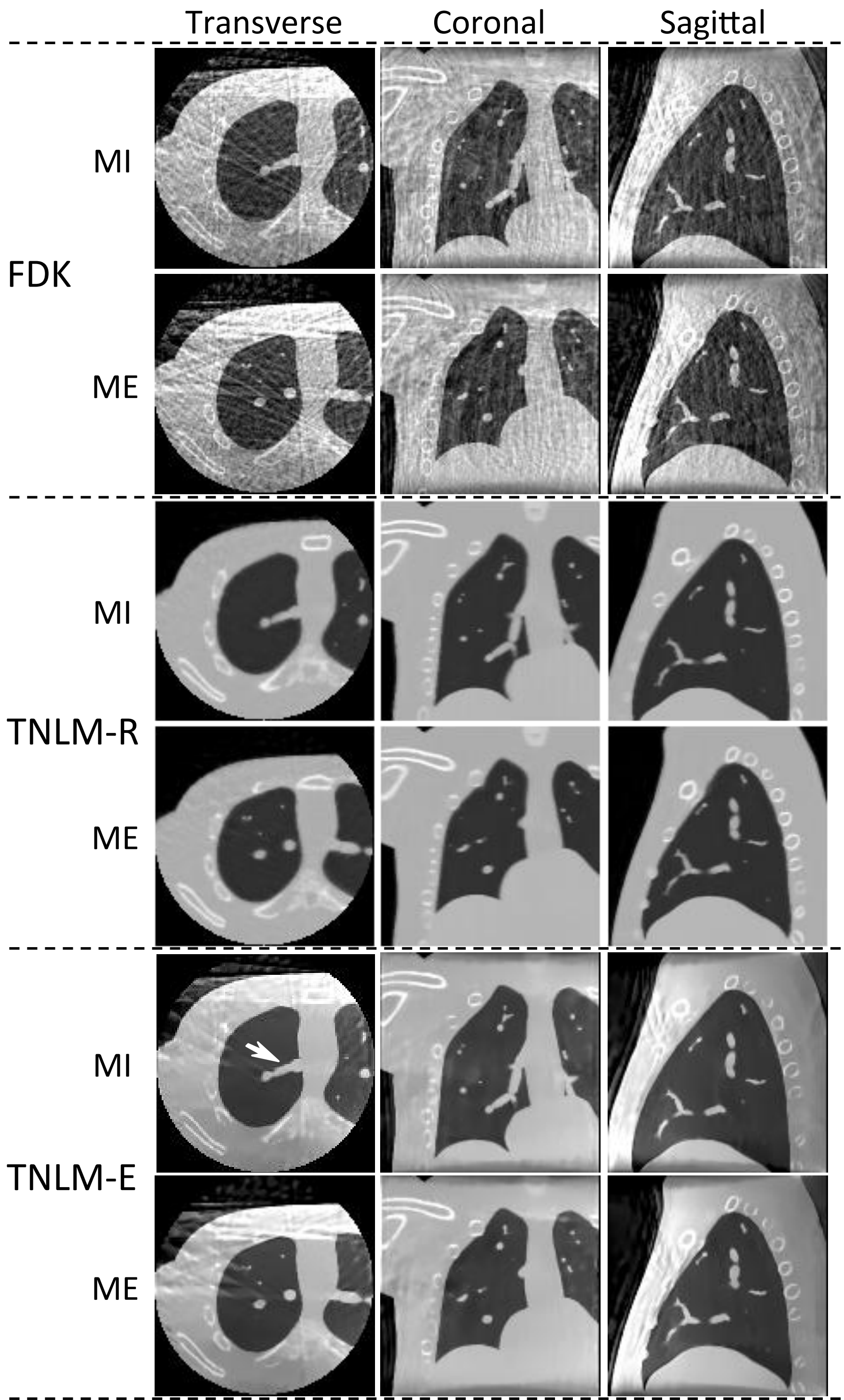

**Figure 2.** 4D-CBCT images of a NCAT phantom at the MI phase and the ME phase. Top two rows: reconstructed from the FDK algorithm. Middle two rows: reconstructed using our TNLM-R algorithm. Bottom two rows: the FDK results enhanced with our TNLM-E algorithm. The white arrows indicate the tumor-like structure used to compute $CNR$.

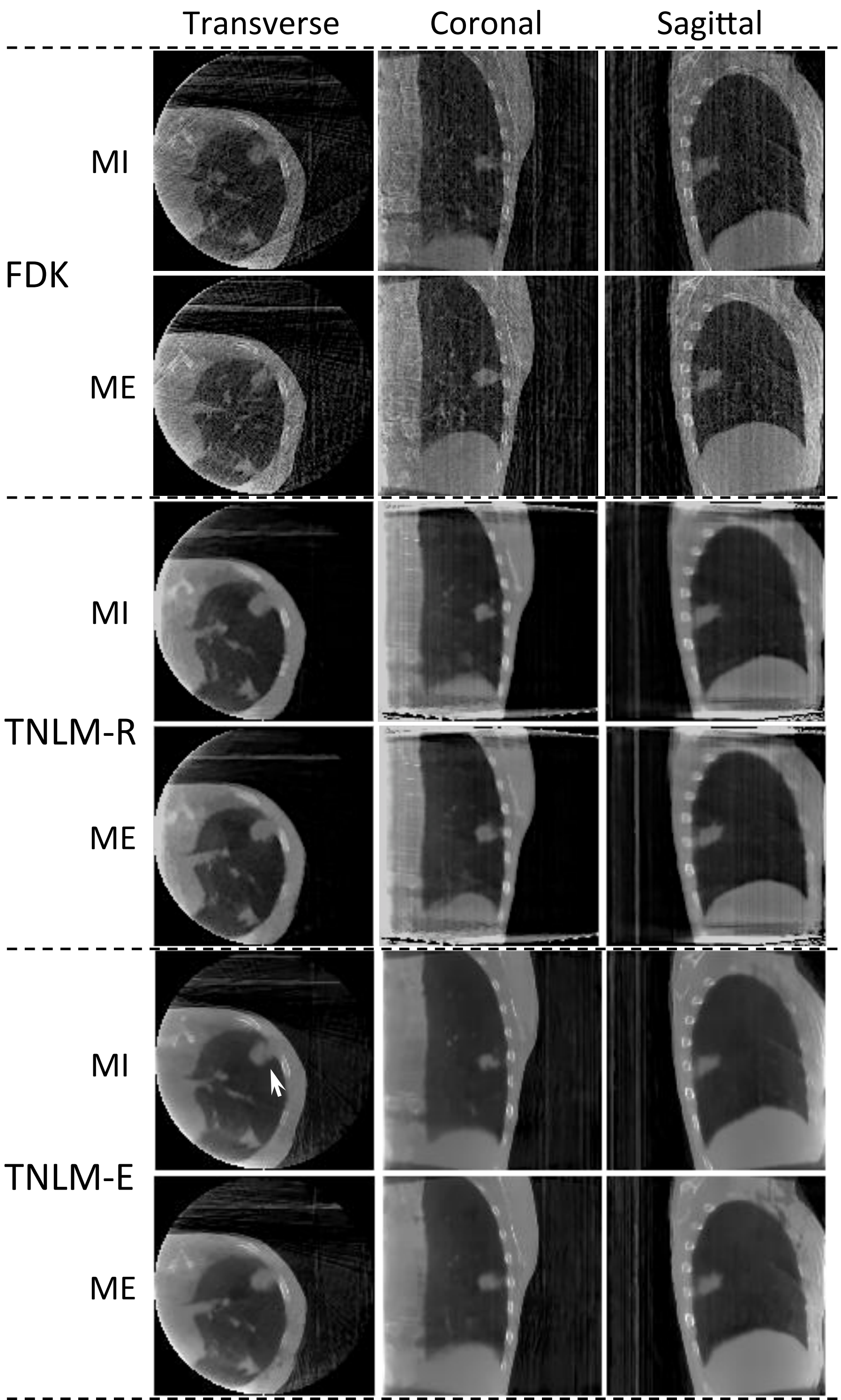

**Figure 3.** 4D-CBCT images of a patient at the MI phase and the ME phase. Top two rows: reconstructed from the FDK algorithm. Middle two rows: reconstructed using our TNLM-R algorithm. Bottom two rows: the FDK results enhanced with our TNLM-E algorithm. The white arrows indicate the tumor-like structure used to compute $CNR$.

**Table 1.** $CNR$s and $SRR$s for the 4D-CBCT images of the NCAT phantom and the patient case obtained using various algorithms.

| | Algorithm | NCAT | Patient |
|---|---|---|---|
| $CNR$ | FDK | 6.8149 | 3.6959 |
| | TNLM-R | 21.3043 | 9.4593 |
| | TNLM-E | 22.7081 | 10.1515 |
| $SRR$ (%) | TNLM-E | 85.09 | 88.27 |

To report the effects of reducing streak artifacts for the entire 4D-CBCT set, we have also computed the $SRR$ for each phase and report the average over 10 phases. As seen in Table 1, 85% and 88% of the streaks in the FDK results are removed by the TNLM-E algorithm, which clearly demonstrates its efficacy.

*3.3 Computational efficiency*

Our TNLM-based 4D-CBCT reconstruction and enhancement algorithms are implemented on NVIDIA CUDA programing environment using an NVIDIA Tesla C1060 card. The total computation time, as well as the time per iteration, is listed in Table 2, where the total computation time is for 7 iterations for the TNLM-R algorithm and 10 iterations for the TNLM-E algorithm, corresponding to the results shown in Fig. 2 and Fig. 3. The time per iteration for TNLM-R algorithm is longer than that in the TNLM-E algorithm. This can be ascribed to the fact that the enhancement algorithm is same as the sub-problem of the reconstruction in terms of computational complexity. Yet, the difference between them is small, indicating that the time spent on solving the sub-problem (P1) in Eq. (5) is short and the majority of computational complexity of our algorithms is from those nonlocal operations.

**Table 2.** Total computation time $t_{tot}$ and time per iteration $t_{iter}$ of our TNLM-based algorithms using an NVIDIA Tesla C1060 GPU card.

| | NCAT (sec) | | Patient (sec) | |
|---|---|---|---|---|
| | $t_{tot}$ | $t_{iter}$ | $t_{tot}$ | $t_{iter}$ |
| TNLM-R | 468.37 | 66.91 | 469.45 | 67.06 |
| TNLM-E | 550.69 | 55.07 | 610.00 | 61.00 |

## 4. Discussions

Since the TNLM-E algorithm takes the FDK reconstruction results as inputs, the resulting image quality highly depends on the input images, specifically, on whether those true anatomical structures can be observed in the FDK results. For the NCAT phantom, due to the limited number of projections in each phase bin, *i.e.* 30, some structures are hardly resolved by the FDK algorithm. For instance, the vertebral body and the sternum are seriously contaminated by the streaks. In this context, the TNLM-E algorithm is not able to find the similar structures between breathing phases, resulting in regions with low image quality. See the area close to the sternum in the last two rows of Fig. 2. While for those areas where clear structures can already be observed in the FDK results, such as inside the lung region of the NCAT phantom, TNLM-E algorithm can distinguish between these structures and the streaks and is able to suppress the streaks to a satisfactory extent. This is also the case for the patient case, where a relatively large number of projections are available in each phase. Although both clear patient anatomy and artifacts are seen on the FDK results, the TNLM-E algorithm effectively suppresses the latter.

In contrast, FDK results are only used to initialize the iteration process in the TNLM-R algorithm. Its performance is not largely related to the corresponding FDK results. For the NCAT phantom, with high quality projections the TNLM-R algorithm is capable of reconstructing images of high quality, demonstrating its advantages over the TNLM-E algorithm. Especially, the vertebral body and the sternum can be clearly observed and overall there are less streak artifacts in the entire images than in the TNLM-E results. When it comes to the patient case, it is found that the results of TNLM-R are not obviously superior to those of TNLM-E, possibly owing to the following two reasons. First, the FDK algorithm performs very well in this case by itself due to the sufficient number of projections and hence provides input images of good quality for the TNLM-E algorithm. Second, TNLM-R algorithm encounters a well-known data truncation issue in this case. In an iterative CBCT reconstruction problem, when the reconstructed region is smaller than the whole patient body, it is quite hard to satisfy projection condition $Pf = y$ in the reduced reconstruction region, leading to some areas around the image boundary where the image intensity tends to blow up. See, for example, the artifacts around the superior and inferior regions of the TNLM-R results in Fig. 2. On the other hand FDK algorithm is a direct algorithm and does not have the truncation issue. The enhancement model therefore does not encounter this problem. Since the main objective here is to study the use of TNLM method in 4D-CBCT, for the NCAT phantom we deliberately truncate the phantom before generating projections and hence the truncation problem does not appear in the TNLM-R algorithm.

Although the computation time for these cases is long and cannot compete with the FDK reconstruction algorithm, the efficiency we have achieved is already a considerable improvement over the CPU implementation. In our work, we did not implement our algorithms on CPU due to the unacceptably long computation time. Since there is no work on the TNLM algorithm reported previously to our knowledge, we estimate its CPU computation time based on NLM algorithm due to the similar computation complexity and algorithm structure. In fact, even for the NLM algorithm, its application is mainly limited to 2D images because of its computational complexity. For those 3D image processing problems with the NLM method, the NLM filter is usually used, in the sense that the image quality is enhanced by using a scheme akin to the Eq. (8) but for only one iteration step. For instance, it has been reported that it takes 21790 sec to perform a NLM operation on a 3D MRI image for an image denoising purpose[43] using a typical CPU. Considering the image size difference ($181\times217\times181$ for this MRI case and $128^3$ for our problem) and the fact that there are 10 phases in our problem, it can be estimated that the computation time for one step TNLM update is about 18 hours on CPU for the cases studied in this paper. Comparing this number with those listed in Table 2, we have achieved a considerable speed up using the GPU card over a typical CPU. Despite the high speedup factor, this long computation time is not yet acceptable for routine clinical application. Other techniques such as further algorithm optimization and multi-GPU implementation may lead to a further efficiency boost, which will be our focus in the future.

## 5. Conclusion

In this paper, we have developed a novel iterative 4D-CBCT reconstruction algorithm and an enhancement algorithm via temporal regularization. The 4D-CBCT images of different phases are reconstructed or enhanced simultaneously by minimizing a whole energy function consisting of a data fidelity term of all the respiratory phases and a temporal regularization between every two neighboring phases, in which a TNLM method is employed to take the temporal redundancy of the 4D-CBCT images into account. The only difference between the reconstruction algorithm and enhancement algorithm is that, in our reconstruction algorithm the data fidelity term is to enforce the consistency between the reconstructed image and the measured projections, while in the other the data fidelity term ensures that the enhanced images does not deviate from the input images largely. The energy functions in these two algorithms are minimized utilizing a forward-backward splitting algorithm and a Gauss-Jacobi update scheme. These algorithms are implemented on a GPU platform to improve their fficiency.

We have tested our algorithms using on a digital NCAT phantom and a clinical patient case. The experimental results indicate that both our reconstruction and enhancement algorithms lead to better image quality than the conventional FDK algorithm. In particular, quantitative evaluations indicate that, compared with the FDK results, our TNLM-R method improves $CNR$ by a factor of 2.56~3.13 and our TNLM-E method increases the $CNR$ by 2.75~3.33 times. The TNLM-E method also removes over ~80% of the streak artifacts from the FDK reconstruction results. The total computation time is up to ~610 sec for the TNLM-R algorithm and ~460 sec for the TNLM-E algorithm on an NVIDIA Tesla C1060 GPU card.

Comparing the TNLM-R and TNLM-E algorithms, it is found that the two algorithms attain their own advantages as well as disadvantages. TNLM-E is slightly better in terms of computation time. Yet, its resulting image quality is limited by the input 4D-CBCT images obtained by other algorithms such as FDK. Especially when there are insufficieny number of projections, 4D-CBCT image quality may not be satisfactory. On the other hand, the TNLM-R algorithm reconstructs images from the x-ray projections directly. In the absence of other problems such as data truncation, the resulting image quality is higher than that from TNLM-E, as evidenced by the NCAT phantom case. The computation time is, however, prolonged due to one more subproblem to solve compared to the TNLM-E algorithm. At present, users may choose one of the two algorithms for an overall consideration of image quality and performance. Our future work will focus on the further improvement of computational efficiency by algorithmic optimization and using hardware with higher performance, as well as to develop better algorithms to improve image quality.

**Acknowledgements**

This work is supported in part by NIH (1R01CA154747-01), Varian Medical Systems through a Master Research Agreement, the Thrasher Research Fund, and the University of California Lab Fees Research Program.